\documentstyle[pra,aps,twocolumn,epsfig]{revtex}
\begin{document}
\newtheorem{lemming}{Lemma}
\draft
\def\uar{\uparrow}
\def\dar{\downarrow}
\def\mmz{{[m_\uar m_\dar 0]}}
\def\vhat{{\hat v}}
\def\rhobar{{\overline\rho}}
\def\sbar{{\overline s}}
\def\fifth{\frac{1}{5}}
\def\uu{{\uar\uar}}
\def\dd{{\dar\dar}}
\def\ud{{\uar\dar}}
\def\du{{\dar\uar}}
\def\izppi{\int\limits_{0^+}^{\infty}}
\def\IB{I_{\rm B}}
\def\gtw{{\tilde{g}}}
\def\ttw{{\tilde{t}}}
\def\thtw{{\tilde{\theta}}}
\def\tot{\frac{3}{2}}
\def\kF{{k_\ssr{F}}}
\def\kJ{{k_\ssr{J}}}
\def\thb{{\bar \theta}}
\def\phb{{\bar \phi}}
\def\Gb{{\overline G}}
\def\II#1#2{I_{#1}^{\ssr[#2\ssr]}}
\def\avg#1{{\overline{[#1]}}}
\def\Var{{\rm Var}}

\twocolumn[\hsize\textwidth\columnwidth\hsize\csname
@twocolumnfalse\endcsname
\title{Wigner Crystalline Edges in 
$\nu \lesssim 1$ Quantum Dots}

\author{
Eyal Goldmann\\
Scot R. Renn }
\address{
Department of Physics, University of California at San Diego,
La Jolla, CA 92093}

\date{\today}

\maketitle

\begin{abstract}
We investigate the edge reconstruction phenomenon believed to occur in 
quantum dots in the quantum Hall regime when the filling 
fraction is $\nu \lesssim 1$.
Our approach involves the examination of  large dots 
($\leq 40$ electrons)
 using a partial diagonalization technique in which the occupancies of the
deep interior orbitals are frozen. To interpret the results of this 
calculation, we evaluate the overlap between the diagonalized ground 
state
and a set of trial wavefunctions which we call  projected necklace (PN) 
states.
A PN state is  simply the angular momentum projection of 
a maximum density droplet surrounded by a ring of localized electrons.
Our calculations reveal that PN  states
have up to 99\% overlap with the diagonalized ground states, and are
lower in energy than the states identified in 
Chamon and Wen's study of the edge reconstruction.
\end{abstract}


\vskip2pc]


\section{Introduction}
A number of recent experiments have studied the behavior of a small 
number $N$ ($1 \sim 100 $) of electrons which have been confined to 
two-dimensional quantum dots.
When these quantum dots are 
subjected to large perpendicular magnetic fields,  a rich phenomenology
may be observed, much of it due to the same physics 
underlying the bulk quantum Hall effect\cite{review,review2}.
In particular, there has been a great deal of interest in the 
maximum density droplet (MDD)
\cite{ashoori,klein,klein2,kouwenhoven,oaknin2,lubin,karlhede,chamon,muller,macdonald,oaknin1,ferconi,ahn,reimann},
which is the small $N$ analog of the bulk $\nu=1$ state.
Experimentally, the MDD has been identified in several studies,
including those of  Ashoori et al.\cite{ashoori} , Klein et
al.\cite{klein,klein2},
and Oosterkamp et al \cite{kouwenhoven}.  
Because the MDD's size decreases monotonically with magnetic field, it is 
clear from electrostatic considerations that a sufficiently strong 
magnetic field must 
destabilize the MDD into a lower density state.  A number of theoretical 
studies have discussed a type of 
instability called an edge reconstruction, 
in which the decrease in charge density first occurs along a ring just 
within the MDD's perimeter.  In this article we demonstrate that, neglecting 
the possibility of spin texturing\cite{oaknin2,lubin,karlhede},
the MDD destabilizes into a state in which 
the droplet edge is best described as a one-dimensional Wigner crystal.   The 
edge reconstructed density profile may then be understood as an angular
momentum projection of the crystallized edge.

In an early study of spin-polarized edge reconstructions, 
Chamon and Wen \cite{chamon} identified approximate ground states of the 
quantum dot by minimizing its energy 
over single Slater determinants of symmetric gauge 
lowest Landau level states.
They found that increasing the magnetic field forces a transition
in which several electrons simultaneously  move from  {\it adjacent}
states in the interior of the droplet to states at the edge.
This state, hereafter referred to as the compact ring (CR) state, therefore
consists of a $\nu=1$  central disk surrounded by a ring in which 
all single particle states are completely occupied.

Other studies on smaller dots have  revealed a rather  
different structure.  For example, Maksym has studied the 
electron-electron correlation functions of small droplets ($N \leq 6$)
using exact diagonalization \cite{maksym}.  
His work finds that for $\nu \ll 1$, the electrons occupy distinct, 
localized positions relative to each other in a structure which 
resembles a small piece of a Wigner crystal\cite{wigner}. 
In a study of somewhat larger dots, ($N \leq 20$),
Muller and Koonin\cite{muller} found that for $\nu \lesssim 1$, 
an approximate Hartree-Fock (HF) Hamiltonian  
is minimized by a state whose 
density profile consists of a $\nu \lesssim 1$
central droplet surrounded by a ring of localized electrons.  

Given that small dots are expected to Wigner crystallize at high magnetic 
fields,  it seems reasonable to expect 
that similar behavior would occur in larger dots.
However, the crystallization of large dots would differ from that of small
dots in at least one important respect: it would occur nonuniformly.
This is most evident if one considers a
droplet which begins as an  MDD, and then evolves into a crystallized state
as the magnetic field increases.
In this case, any portion of the dot which crystallizes must be partially 
depopulated, 
and the extra charge moved to orbitals beyond the MDD perimeter.
Since this transfer of charge costs less confinement energy for electrons
which begin near the droplet edge, we expect the droplet edge to crystallize
first, while the droplet center initially remains liquid\cite{reimann1}.

In  the following discussion, we will present numerical evidence that 
the MDD does, in fact, destabilize into an edge crystallized state 
when the magnetic field increases.
To be precise, we will present a partial diagonalization of
 the many body Hamiltonian with the deep interior orbitals frozen.
We will then calculate the overlap of this ground state with  the 
``projected necklace'' (PN), a trial wavefunction which
 exhibits  edge crystallization. See Fig. (2).  
As a result, we find  that PN  states have high overlap with 
diagonalized ground states, and have lower energy than 
Chamon and Wen's compact ring (CR) state.
In addition, we have found that a reconstruction to a PN 
state may exhibit a much weaker addition spectrum signature
than does a reconstruction to a CR state.
Whether the reconstruction signature is, in fact, weak depends on the 
detailed form of the confinement potential.

\section{Projected Wigner Necklace States}
We begin our study by considering
parabolically confined two-dimensional electrons in the presence of a 
perpendicular 
magnetic field.  The noninteracting Hamiltonian for this problem is
\begin{equation}
H_0=\frac{1}{2m^*}({\bf p} + \frac{e}{c}{\bf A})^2  +
    \frac{1}{2}m^* \omega_0^2 r^2,
\label{H0}\end{equation}
where $m^*$ is the band mass, $\omega_0$ is the confinement strength 
(typically $\hbar \omega_0 \sim$ 1--3 meV), and 
${\bf A} = \frac{B}{2}({\bf {\hat z}}
\times{\bf r})$.
For sufficiently high $B$, the only appreciably occupied eigenstates
of $H_0$ are the wavefunctions of the lowest Fock-Darwin level (LFDL)
\cite{review},
\begin{equation}
\psi_l({\bf r}) = (2^{l+1}\pi l!)^{-1/2} (\frac{z}{l_0})^l 
                   e^{-|z|^2/4l_0^2},
\label{psi}\end{equation}
where $l$ is the angular momentum, $z=x-iy$, and 
$l_0$ is the characteristic length, given by 
$l_0 = l_m \kappa^{-1}$, with $l_m = (\frac{\hbar c}{e B})^{1/2}$ ,
$\kappa = (1 + 4\frac{\omega_0^2}{\omega_c^2})^{1/4}$, and
$\omega_c = \frac{eB}{m^* c}$.
The angular momentum of $\psi_l$ is $l$, and its energy is 
\begin{equation}
\epsilon_l = \frac{1}{2} \hbar \omega_c 
\left( \kappa^2 \left(l+1 \right) -l \right).
\label{epsilon}\end{equation}
Using the $\psi_l$'s as our basis states, we introduce  
the second-quantized Hamiltonian for many-electron quantum dots, 
\begin{equation}
H = \sum_l \epsilon_l c_l^\dagger c_l +
\frac{1}{2}\sum_{ijmn} V_{ijmn}c_i^\dagger c_j^\dagger c_n c_m.
\label{H}\end{equation}
Here,  $V_{ijmn}$ is the Coulomb interaction\cite{stone}
\begin{equation}
V_{ijmn} = \frac{e^2}{\epsilon}
           \int d^2 r_1 d^2 r_2 \psi^*_i({\bf r_1}) \psi^*_j({\bf r_2})
           \frac{1}{|{\bf r_1} - {\bf r_2}|}
           \psi_m({\bf r_1}) \psi_n({\bf r_2}), 
\label{Coulomb}\end{equation}
and $c_l^\dagger (c_l)$ is the creation (annihilation) operator 
corresponding to $\psi_l$.

For the special case of parabolic confinement, the eigenstates of $H$ depend 
on $B$ only through the $l_0$ dependence of the single particle orbitals.
When $B$ increases, $l_0$ decreases, and the many-electron states 
squeeze inward, thereby increasing their Coulomb energy. 
It follows that, when a quantum dot is subjected to a steadily increasing
magnetic field, its general inward motion will be punctuated by a series of
transitions in which the dot jumps to ground states of larger area,
 and accordingly, larger  angular momentum,
in order to decrease its Coulomb energy.

The state of lowest 
total angular momentum $L$ entirely contained in the spin-polarized LFDL
is the MDD\cite{chamon,macdonald},
which is simply a single Slater determinant of the $N$ LFDL 
single particle states with angular 
momenta $0,1,\dots,N-1$.
Consequently, when $N$ is small,
the MDD is the first state in  the spin-polarized LFDL
to become a ground state of the quantum dot as $B$ increases from zero.
A further increase in $B$ will compress the MDD, 
so that it eventually becomes smaller in diameter 
than the  classical electrostatic solution
for a parabolically confined 2D droplet,
which is a hemisphere \cite{electrostatics}.
At this point the contrast between the electronic density and the  
electrostatic solution is most pronounced at the droplet edge: 
The electronic density slightly 
inside (outside)  the MDD edge 
is greater (less)  than would be expected from the electrostatic solution. 
Consequently, the initial destabilization  of the MDD at high $B$ involves a 
redistribution of the charge density at the  MDD edge, 
and leaves the $\nu=1$ center intact
\cite{previous}. 

We seek to characterize these instabilities by introducing a set of 
trial wavefunctions which incorporate the possibility of crystalline 
correlations at the $\nu=1$ edge.
To begin, we construct a state which consists of a compact central disk
of $D$
electrons surrounded by a necklace of $R=N-D$ localized
 electrons equally spaced along a ring of radius $u$.
Electrons in the central disk are assumed to 
occupy wavefunctions $\psi_l$, given by Eq. (\ref{psi}). Ring electrons are 
assumed to have orbitals
\begin{equation}
\chi_p({\bf r}) = (2 \pi)^{-1/2} 
\exp \left[ - \frac{1}{4l_0^2} \left( |{\bf r} - {\bf u_p}|^2 
                                    - 2i {\bf \hat{z} \cdot r \times  u_p}
     \right) \right]
\label{chi}\end{equation}
where ${\bf u_p} = u(\cos \phi_p,\sin \phi_p)$ are the guiding centers, and 
$\phi_p = \frac{2 \pi p}{R}$, with $p = 0,\dots,R-1$.
Our many-body wavefunction is thus
\begin{equation}
\langle {\bf r}_0,\dots,{\bf r}_{N-1} \vert {N,R,u;WN} \rangle = 
{\mathcal A} \left(
             \prod_{k=0}^{D-1} \psi_k({\bf r_{\it k}}) 
            \prod_{l=D}^{N-1} \chi_{l-D}({\bf r_{\it l}}) 
             \right) 
\label{necklace}\end{equation}
where ${\mathcal A}$ is the antisymmetrization operator.  
We call this state a ``Wigner necklace''.  The density profile of a  
Wigner necklace with $N=40$, $R=18$, and $u=10.5l_0$ is plotted in Fig. (1).
\begin{figure} 
\centering
\leavevmode
\psfig{figure=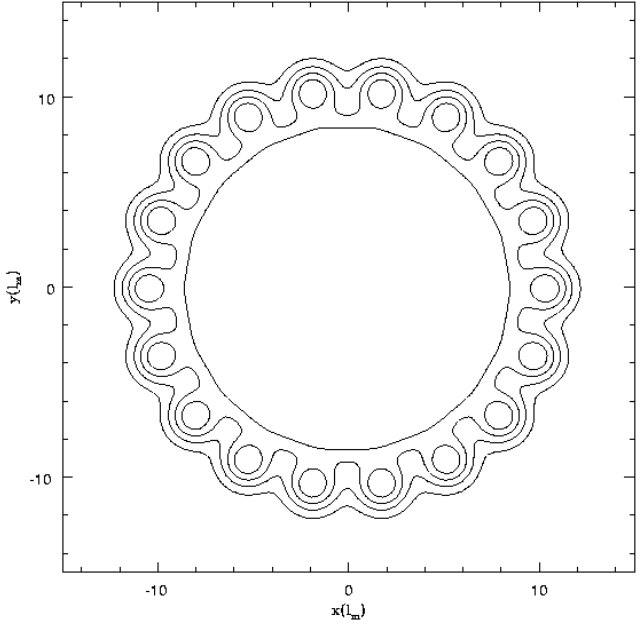,width=\linewidth}
\caption{Density contours of a Wigner necklace state with $N=40$, $R=18$,
and $u=10.5l_0$.}
\label{figure1}
\end{figure}

In order to project
$ \vert {N,R,u;WN} \rangle$ onto fixed total angular momentum $L$,
we seek to expand it into a sum over Slater determinants of the 
$\psi_l$'s.
Because only states with a $D$ electron
compact central disk contribute to the expansion,
it is convenient to define $\vert D, m_1,\dots,m_{R} \rangle$
as an $N$ body state
consisting of a $D$ electron compact central disk and $R$ additional 
single particle orbitals, that is,
\begin{eqnarray}
\langle {\bf r}_0, \dots, {\bf r}_{N-1} &\vert& D, m_1,\dots,m_{R} \rangle = 
\nonumber \\
&{\mathcal A}& \left(
            \prod_{k=0}^{D-1} \psi_k({\bf r_{\it k}})
            \prod_{l=D}^{N-1} \psi_{m_{l-D+1}}({\bf r_{\it l}}) 
             \right).
\end{eqnarray}
Defining $P(L)$ to be the operator which
projects onto total angular momentum $L$, we thus have 
\begin{eqnarray}
P(L)\vert N,R,u;WN \rangle = 
\sum_{m_1+\dots+m_{R}=L-\frac{1}{2}D(D-1)}  \nonumber \\
\vert D, m_1,\dots,m_{R} \rangle
\langle D, m_1,\dots,m_{R} \vert N,R,u;WN \rangle, 
\label{expand}\end{eqnarray}
where the sum only includes Slater 
determinants with the correct total angular momentum.  
In practice, it is necessary to truncate this sum by disposing of all Slater 
determinants involving single particle momenta with $m_i > N+k-1+W $,
where $W$ acts as a cutoff.
In this work, we use $W \geq 7$.
The coefficients in Eq. (\ref{expand}), 
which are computed in the Appendix, are given by 
\begin{eqnarray}
\langle D, m_1,\dots,m_R \vert N,R,u;WN \rangle  =  \nonumber \\
C
\prod_{l=1}^R (m_l!)^{-1/2} 
\prod_{1\leq i<j \leq R} \sin \biggl(\frac{\pi}{R} (m_i-m_j)\biggr)
\label{coef_result} \end{eqnarray}
where $C$ is a constant common to every coefficient in the expansion of a 
particular $\vert N,R,u;WN \rangle$.

It follows from (\ref{coef_result}) (see the Appendix) 
that if we define $P(L)$ to be the operator which
projects states onto total angular momentum $L$, then 
$P(L) \vert N,R,u;WN \rangle = 0$ unless $L = \frac{1}{2} N(N-1) + kR$, 
where $k=0,1,\dots$ (A similar result has been 
derived in Refs. \cite{maksym}, \cite{magic2}, and \cite{magic1}
for the case of totally crystallized droplets.).
Hence, the angular momentum projected states are
conveniently defined as 
\begin{equation}
\vert N,R,k;PN \rangle = P(\frac{1}{2} N(N-1) + kR) \vert N,R,u;WN \rangle,
\label{project}\end{equation}
where PN means ``projected necklace.'' Note that $\vert N,R,k;PN \rangle$
is independent of the ring radius $u$ used to construct the unprojected 
$\vert N,R,u;WN \rangle$.  The $k=1$ states are particle-hole excitations of 
the MDD\cite{oaknin1}.

To illustrate the effect of the angular 
momentum projection,  we examine the electron-electron correlation 
function
\begin{equation}
P({\bf r},{\bf r_1}) = \langle 
\Psi^\dagger({\bf r}) \Psi^\dagger({\bf r_1}) \Psi({\bf r_1}) \Psi({\bf r})
\rangle ,  
\label{correlation}\end{equation}
where $ \Psi({\bf r})= \sum_{l=0}^\infty \psi_l({\bf r}) c_l$
\cite{maksym,hausler}.
In Fig. (2), we plot $P({\bf r},{\bf r_1})$ for the state 
$\vert 40,18,3,PN \rangle$, where we have fixed
${\bf r_1} = (x,y) = (7.5 l_0,0)$.
\begin{figure} 
\leavevmode
\psfig{figure=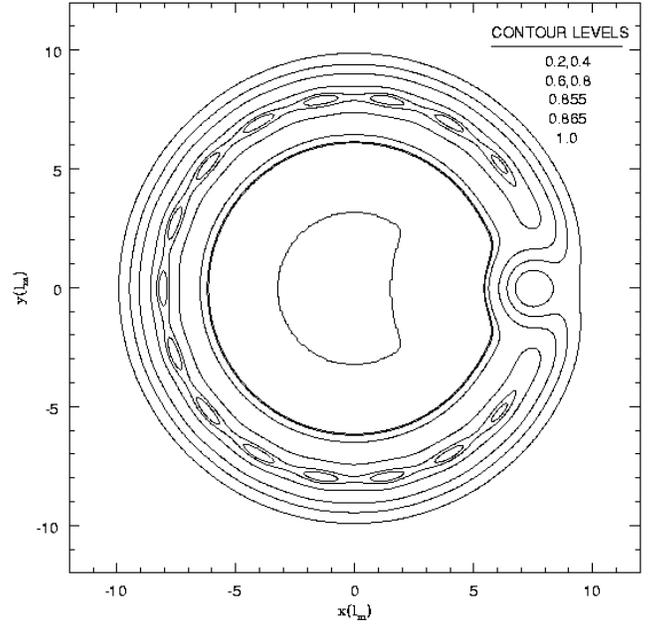,width=\linewidth}
\caption{Two particle correlation function $P({\bf r},{\bf r_1})$
for the state $\vert 40,18,3;PN \rangle$,
plotted as a function of ${\bf r}$, with ${\bf r_1}$
fixed at ${\bf r_1} = (7.5l_0,0)$.
$P({\bf r},{\bf r_1})$ was rescaled so that its maximum value is one.
In order to emphasize the state's
axial modulation, extra contours were added at levels  0.855 and 0.865.
}
\label{figure2}
\end{figure}
We note that this state is the lowest energy PN state with $N=40$ 
when $\hbar \omega =1.6{\rm meV}$ and $B=2.42 {\rm T}.$ See Fig. (6) and 
the accompanying discussion, which will follow.
Indeed, a modulation of $P({\bf r},{\bf r_1})$ is evident at 
the droplet's edge, but is much weaker than the localization in the
unprojected WN.   Extra contours were added to the figure in order to
reveal the effect (see the caption). 

It is instructive to note the similarities between the PN states and the 
compact ring (CR) states discussed earlier.  
Formally, we define the CR state $\vert N,R,k;CR \rangle$ to be a 
Slater determinant constructed from the single particle LFDL 
states with angular momenta $l=0,\dots,D-1$ and angular momenta
$l=D+k,\dots,N+k-1$, where $D=N-R$, as above.  
Both $\vert N,R,k;CR \rangle$ and $\vert N,R,k;PN \rangle$
may be thought of as states formed by moving $k$ electrons 
from the bulk to the edge of an $N$ electron MDD in order to separate an 
$R$ electron ring from an $N-R$ electron core.
Furthermore, for a given triplet $(N,R,k)$, 
$\vert N,R,k;CR \rangle$ and $\vert N,R,k;PN \rangle$
have the same total angular momentum $L$.   It follows from 
Eqs. (\ref{epsilon}) and (\ref{H}) that 
the sum of kinetic and confinement energies of the two states is the 
same for all $B$.  Therefore, the state with the lower Coulomb energy 
will be favored for all $B$.

Motivated by this observation, we have made pairwise comparisons of the 
Coulomb energies of the states
$\vert N,R,k;CR \rangle$ and $\vert N,R,k;PN \rangle$
for all $(N,R,k)$ such that $25 \leq N \leq 60$, $2 \leq R \leq 25$,
and $1 \leq k \leq 10$.  
The ranges of $R$ and $k$ have been selected to include all PN 
and CR states 
which are likely to become ground states of the quantum dot at magnetic 
fields slightly beyond the stability range of the MDD.
We have confirmed this selection
with appropriate minimizations of the droplet energy with respect to PN states,
and in separate calculations, with respect to CR states.

For nearly each triplet $(N,R,k)$ within the stated range, we find that 
$\vert N,R,k;PN \rangle$ has the lower Coulomb energy, 
although the difference is small, typically around $0.1\% \sim 0.3\%$.  
A sample of this comparison is given in Table (1).
\begin{table} 
\large
\begin{center}
\begin{tabular}{|c|c|c|c|} \hline
$(N,R,k)$
 & \shortstack{Compact\\Ring}
 & \shortstack{Projected\\Necklace}
 & $E(CR) - E(PN)$\\ \hline \hline

(40,18,1) & 124.577 & 124.487 & 0.090 \\ \hline
(40,18,2) & 123.350 & 123.222 & 0.128 \\ \hline
(40,18,3) & 122.123 & 121.982 & 0.141 \\ \hline
(40,18,4) & 120.914 & 120.771 & 0.143 \\ \hline
(40,18,5) & 119.732 & 119.593 & 0.139 \\ \hline
(40,18,6) & 118.585 & 118.450 & 0.134 \\ \hline
(40,18,7) & 117.475 & 117.345 & 0.131 \\ \hline
(40,18,8) & 116.406 & 116.277 & 0.129 \\ \hline
(40,18,9) & 115.377 & 115.247 & 0.130 \\ \hline
(40,18,10)& 114.389 & 114.260 & 0.129 \\ \hline
\end{tabular}
\end{center}
\caption[Comparison of the Coulomb energies of compact ring states
and projected necklace states]
{Comparison of the Coulomb energies of compact ring states
and projected necklace states with $N=40$, $R=18$, and $k=3$.
All energies are in units of
$\frac{e^2}{\epsilon l_0}$.}
\label{table_energies}
\end{table}
The only triplets in the stated range
for which the corresponding CR state has the lower Coulomb energy are 
$(N,2,10)$ for $N=25,\dots,30$ and  $(25,3,10)$.  However, these 
exceptions are not a concern because in this range of $N$, $R \gg 3$
for the ground state, regardless of whether 
one assumes that the ground states are CR states 
or PN states.    
These results are 
encouraging, as they show that in the vicinity of the edge reconstruction, 
the ground state of a HF calculation
will necessarily be unstable with respect to at least one PN state.

\section{Diagonalization Calculations}

We wish to further test
the validity of the PN states by comparing them with the 
results of diagonalization calculations.  Unfortunately, the systems  we 
are studying are too large to accommodate complete diagonalization.
Instead, we restrict the basis of our diagonalization calculation to 
Slater determinants of LFDL states which include a compact central disk
of $D$ electrons, defining $\vert N,R,k;GS\rangle$ to be the lowest 
energy eigenstate of $H$ within this basis for which $L = L_{MDD} + kR$,
where $L_{MDD} = \frac{1}{2}N(N-1)$. This restriction
is reasonable because, as discussed above, ground states with 
$L \gtrsim L_{MDD}$ are expected to retain a $\nu \approx 1$ center.

In Fig. (3), we display the single particle occupancies of the states 
$\vert 40,18,3,PN \rangle$, $\vert 40,18,3,CR \rangle$, 
and $\vert 40,18,3,GS \rangle$.   
\begin{figure} 
\centering
\leavevmode
\psfig{figure=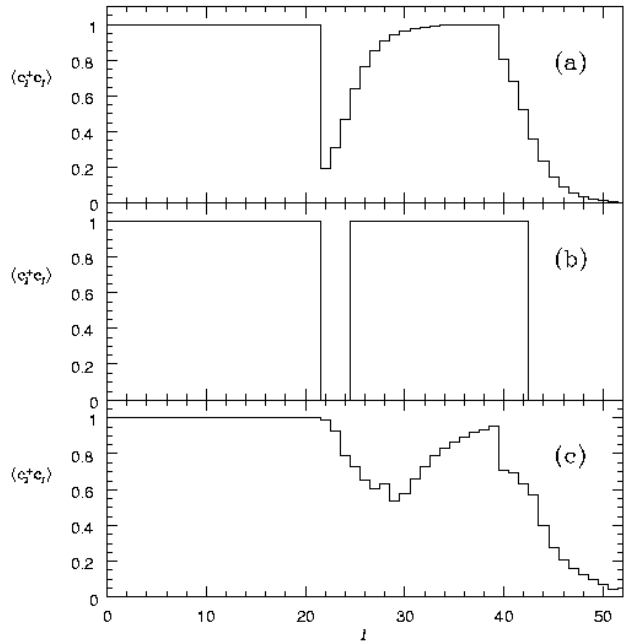,width=\linewidth}
\caption{Panels (a), (b), and (c) show, respectively, the single particle
occupancies of the projected necklace state $\vert 40,18,3;PN \rangle$,
compact ring state $\vert 40,18,3;CR \rangle$, and
diagonalized state $\vert 40,18,3;GS \rangle$.}
\label{figure3}
\end{figure}
Clearly, the PN state resembles the 
diagonalized state more closely than the CR state does.  The main 
difference  between the PN state and the diagonalized state is that the 
occupancy of the PN state drops sharply at the edge of the central 
disk.  This discrepancy becomes more pronounced as the 
separation of the ring and central disk increases.

More quantitatively,  we have computed 
$\langle N,R,k;PN \vert N,R,k; GS\rangle$, the overlap between the 
diagonalized ground state and the projected necklace state, 
for states with $N=30$ and $N=40$.
The results are plotted in Fig. (4).
\begin{figure} 
\leavevmode
\psfig{figure=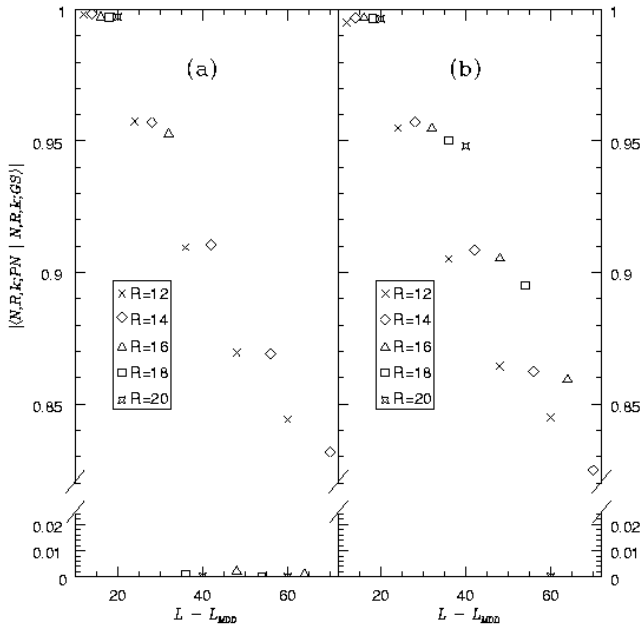,width=\linewidth}
\caption{Overlaps of projected necklace states with ground states of
diagonalization calculations. In panels (a) and (b), $N=30$ and
$N=40$ electrons, respectively.  The
data in both panels has the pattern of descending staircases, with each
step corresponding to a single value of $k$.}
\label{figure4}
\end{figure}
Interestingly, $\langle N,R,k;PN \vert N,R,k; GS\rangle$ is 
determined primarily by $k$. This is evident in Fig. (4), 
where the data points 
assemble into a staircase of neat horizontal bunches which descend with 
increasing $k$.  The numerical values of the overlaps presented 
range
between 0.8--1.0 with the higher overlaps corresponding to smaller $k$. 
The $k=1$ states, whose overlaps consistently exceed $0.99$, are
particularly successful \cite{oaknin1}.
On the other hand, 
$\langle N,R,k;PN \vert N,R,k; GS\rangle$ often drops to nearly zero 
(i.e. $\leq 0.002$) with
increasing $k$. 
This effect occurs when some of the ring electrons 
in the diagonalized ground state become adjoined to the central disk, so 
that $\vert N,R,k;GS \rangle$ has, in  effect, fewer electrons in its 
outer ring than 
$\vert N,R,k; PN\rangle$.
As $N$ increases, the preferred number of electrons in the outer ring 
increases,  and  this effect disappears for  
fixed $(R,k)$.
These results, along with the energetic studies discussed above, demonstrate
that the PN states with small $k$ are good 
approximations to the edge reconstructed states.

We have also computed $\langle N,R,k;CR \vert N,R,k; GS\rangle$, the overlap
of a  compact ring state with the corresponding diagonalized state,  for 
the same triplets $(N,R,k)$ for which we computed
$\langle N,R,k;PN \vert N,R,k; GS\rangle$.   
The results are plotted in Fig. (5).
\begin{figure} 
\centering
\leavevmode
\psfig{figure=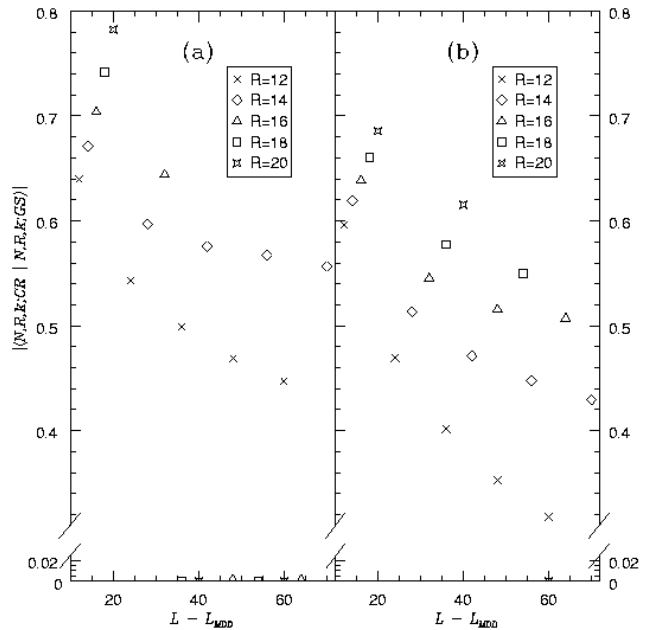,width=\linewidth}
\caption{Overlaps of compact ring (CR) states with ground states of
diagonalization calculations.  In panels (a) and (b) $N=30$ and $N=40$
electrons, respectively.  The CR states are less accurate for the larger
$N$, and are  consistently less accurate than the projected necklace states.
}
\label{figure5}
\end{figure}
The computed values of $\langle N,R,k;CR \vert N,R,k; GS\rangle$  
fall in the range 0.3--0.8, 
except for a few values 
which are nearly zero (i.e. $\leq 0.002$).  
The set of triplets $(N,R,k)$ for which 
$\langle N,R,k;CR \vert N,R,k; GS\rangle$ is nearly zero is exactly 
the same set of triplets for which 
$\langle N,R,k;PN \vert N,R,k; GS\rangle$ is nearly zero.  
As for the majority of triplets, for which the overlap is substantial 
($\geq 0.3$), $\langle N,R,k;CR \vert N,R,k; GS\rangle$ is on 
average  36\% less than the corresponding  PN overlap 
when $N=30$, and 44\% less than the corresponding PN overlap when 
$N=40$.

\section{Addition Spectrum Calculations}
We now discuss the consequences of the PN states for an important experimental
quantity, the quantum dot addition spectrum.
An addition spectrum consists of a sequence of measurements of the
chemical potential
\begin{equation}
\mu(N) \equiv E_0(N) - E_0(N-1), 
\label{mu}\end{equation}
over some range of $N$, 
where $E_0(N)$ is the ground state energy
of an $N$-electron quantum dot
\cite{ashoori,klein,klein2,kouwenhoven}.
To perform this computation, we assume that the ground state 
in the $\nu \lesssim 1$ regime 
is the PN state which minimizes the energy, that is, 
we take 
\begin{equation}
E_0(N) = \min_{R,k} \space \langle N,R,k;PN \vert H \vert N,R,k;PN \rangle.
\label{E0}\end{equation}
The $B$ dependence of $\mu(N)$ for a parabolically confined droplet 
with $\hbar \omega = 1.6 {\rm meV}$, 
as computed from (\ref{mu}) and (\ref{E0}),
is plotted in Fig. (6) for $N=40$ to $N=46$.
\begin{figure} 
\centering
\leavevmode
\psfig{figure=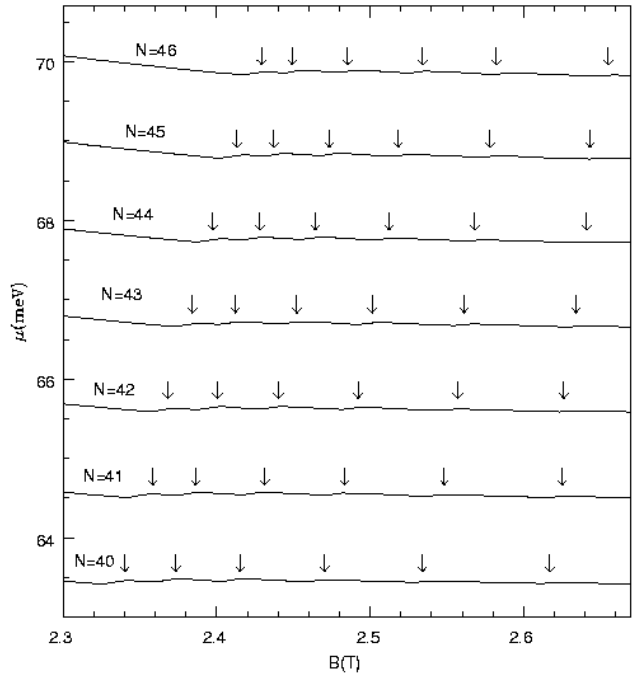,width=\linewidth}
\caption{Addition spectrum  traces for a droplet subject to a parabolic
confinement of strength $\hbar \omega_0  = 1.6 {\rm meV}$, computed
by energy minimization over projected necklace states.  The $N$th line
(as marked) corresponds to the addition of the $N$th electron into
the droplet.  Each arrow indicates a unit  increase
in the $N$ electron droplet's
ground state value of $k$.  The leftmost column of arrows
denotes the edge reconstruction.}
\label{figure6}
\end{figure}

Notably, the addition spectrum 
traces are nearly featureless in the vicinity of the 
edge reconstruction, except for small kinks due to increments in $k$.  
Recall that $k$ may be understood as the number of holes separating the 
central disk and the compact edge.
As $B$ increases,  the ground state value of $k$ for a 
given $N$ always increases monotonically in unit increments.  
The first 
increment of $k$ for a given $N$, from $k=0$ to $k=1$, initiates  the  
edge reconstruction.   
Each increment
in the ground state value of $k$ in the $N$ electron dot is indicated 
in Fig. (6) by an arrow pointing to the $N$th addition line.  
The ground state value of $R$  at the onset of the edge reconstruction
is determined mainly by the requirement that the filling fraction at the center
of the reconstructed edge be  $\nu \lesssim 1$.  As $N$ increases from 
$N=30$ to $N=49$, this value increases from $R=16$ to $R=20$. 

These results are in distinct contrast with the 
HF result of Ref. \cite{chamon}, 
in which the onset of the reconstruction involves 
the simultaneous transfer of several electrons, 
 and causes a sharp upward jump in the addition spectrum. 
However, because our approach is based on a more accurate solution of the 
same Hamiltonian (see preceding discussion), we believe that the cusps seen in 
Ref. \cite{chamon} are an artifact of the HF approximation. 

On the other hand, experimental work has demonstrated that the high $B$ 
destabilization of  the MDD causes distinct cusps in the addition spectrum
of the quantum dot\cite{klein,kouwenhoven}. 
In light of the results presented in Fig. (6), it is thus 
reasonable to ask whether circumstances exist in which the experimental 
observation of addition spectrum cusps would even be consistent with 
spin polarized edge reconstructions.

In fact, we may reintroduce edge reconstruction
induced cusps into the addition spectrum 
by introducing a 
non-parabolic term 
into the confinement potential.  To be specific, we write $V(r)$ as the 
sum of the parabolic confinement potential and an additional coffee cup 
shaped potential, giving 
\begin{equation}
V(r) = \frac{1}{2} m^* \omega_0^2 r^2 + \beta (r-a) \theta (r-a),
\label{coffee}\end {equation}
where $\beta$ and $a$ are constants, and 
$\theta(x)$ is the Heaviside step function.
The second quantized many-body Hamiltonian now becomes
\begin{equation}
H = \sum_l (\epsilon_l + \gamma_l)c_l^\dagger c_l + 
\sum_{ijmn} V_{ijmn}c_i^\dagger c_j^\dagger c_n c_m, 
\label{H_coffee}\end{equation}
with 
$\gamma_l = \beta \int_a^\infty d^2 r \psi_l^*({\bf r}) 
            (r-a) \psi_l ({\bf r}).$

By exact diagonalization of Eq. (\ref{H_coffee})
within the LFDL, we compute the addition spectra of dots
subject to and not subject to a coffee-cup confinement,  and plot the results, 
respectively, in Figs. (7) and (8).
\vspace{1in}
\begin{figure} 
\centering
\leavevmode
\psfig{figure=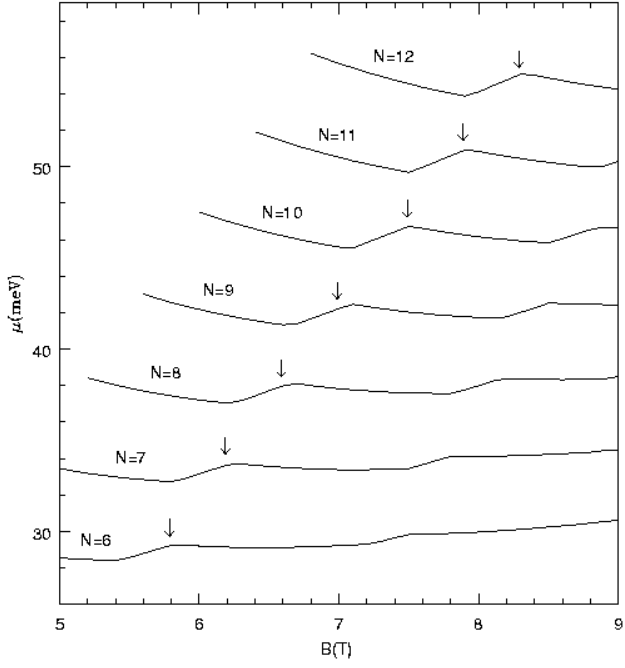,width=\linewidth}
\caption{Addition spectrum traces, computed with exact diagonalization,
for a  droplet subject to
both a parabolic confinement potential  of
strength $\hbar \omega_0 = 3.0 {\rm meV}$, and  a
nonparabolic
confinement potential characterized by parameters
$a = 40{\rm nm}$, and $\beta = 2.0 {\rm meV/nm}$  (See Eq. (\ref{coffee})).
Compare with Fig. \ref{figure8}.}
\label{figure7}
\end{figure}
\begin{figure} 
\centering
\leavevmode
\psfig{figure=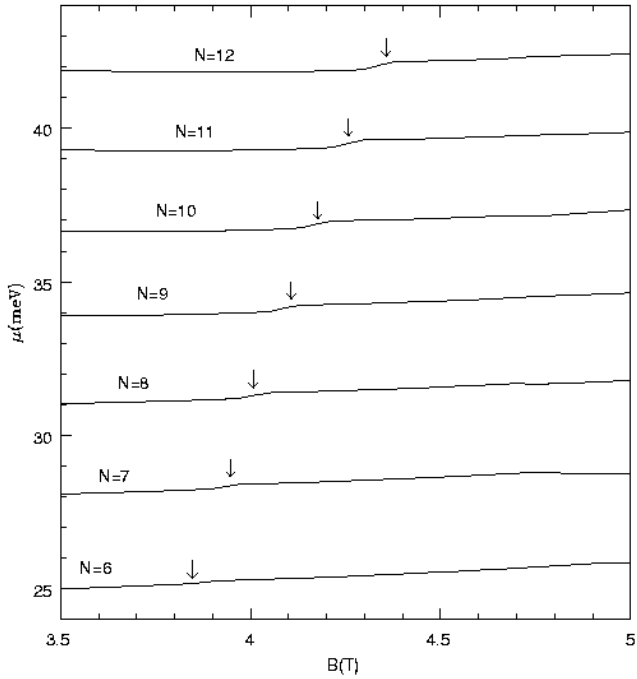,width=\linewidth}
\caption{ Addition spectrum  traces, computed with exact diagonalization,
for  a droplet subject to a parabolic confinement of strength
 $\hbar \omega_0 = 3.0 {\rm meV}$. Compare with Fig. \ref{figure7}. }
\label{figure8}
\end{figure}
For both dots, 
$\hbar \omega_0 = 3.0 {\rm meV}$, and for the dot of Fig. (7), 
$a=40 {\rm nm}$ and $\beta = 2.0 {\rm meV/nm}$.
In these plots, a lone arrow near each addition spectrum trace marks the 
edge reconstruction.  
Indeed, the edge reconstruction
induced kinks of the nonparabolically confined system
are much larger than those of the parabolically confined system.  The 
kink size also increases with electron number, similar to the data of 
Oosterkamp et al \cite{kouwenhoven}.

In summary, we have shown
that projected necklace (PN) states accurately describe the  
spin-polarized instabilities of $\nu \lesssim 1$ quantum Hall droplets.  
PN states whose total angular momentum differs from that of the MDD by 
$70\hbar$ or less typically have overlaps with diagonalized states 
in the range 0.8--0.99.
By comparison, the compact ring (CR) states, which are a generalization of the 
states 
identified by Chamon and Wen \cite{chamon},  have overlaps of 
0.3--0.8 with diagonalized 
states.  We have also shown that near the edge reconstruction, the lowest 
energy PN state is always lower in energy than the lowest energy CR state.
By performing energy minimizations over 
PN states, we have shown that edge reconstruction induced cusps in 
the addition spectra of parabolically confined dots
 are much smaller than suggested by Hartree-Fock \cite{chamon} studies. 
Finally, we have shown that large cusps in the addition spectra may 
occur in dots with  nonparabolic confinement potentials.

\section{Acknowledgements}

We would  like acknowledge support from the Hellman
foundation. We would also like to thank
D.P. Arovas, R.C. Ashoori,
M.D. Johnson, and J.H. Oaknin for useful conversations and
communications.

\appendix
\section*{}

In this appendix, we first compute the coefficient 
$\langle D, m_1,\dots,m_{R} \vert N,R,u;WN \rangle$,
introduced in Eq. (\ref{expand}).
Recall that in general, the overlap of two Slater determinants 
$\vert a_1,\dots,a_K \rangle $ and $\vert b_1,\dots,b_K \rangle$,
where the $a_i$ and $b_i$ are single particle states, is  

\begin{equation}
\langle a_1,\dots,a_K \vert b_1,\dots,b_K\rangle  = 
\left| \begin{array}{ccc}
\langle a_1 \vert b_1 \rangle & \dots  & \langle a_K \vert b_1 \rangle \\
\vdots                        & \ddots & \vdots                        \\
\langle a_1 \vert b_K \rangle & \dots & \langle a_K \vert b_K \rangle 
\end{array}
\right|
\label{blocks}\end{equation}
Hence 
\begin{equation}
\langle D, m_1,\dots,m_{R} \vert N,R,u;WN \rangle = 
{\bf
\left| \begin{array}{ll}
{\bf 1_{\it D}} & {\bf 0_{\it D \times R}} \\
{\bf 0_{\it R \times D}} & {\bf S_{\it R}}
\end{array}
\right|
}
\label{coef}\end{equation}
where ${\bf 1_{\it D}}$  \normalsize 
is the $D \times D$ unit
matrix, due to the inclusion of the $D$ electron compact disk in each state,  
${\bf 0_{\it m \times n}}$ \normalsize 
is the $m \times n$ zero matrix,
and 
\begin{equation} 
{\bf S_{\it R}} = 
{\it 
\normalsize
\left( \begin{array}{ccc}
\langle \psi_{m_1} \vert \chi_0 \rangle & \dots  & \langle \psi_{m_{R}}\vert \chi_0 \rangle \\
\vdots                        & \ddots & \vdots                        \\
\langle \psi_{m_1} \vert \chi_{R-1}\rangle & \dots & \langle \psi_{m_{R}} \vert \chi_{R-1} \rangle
\end{array}.
\right)
}
\end{equation}
\normalsize
To compute ${\bf S}_{\it R}$, we use the result
$\langle \psi_l \vert \chi_p \rangle = \alpha_l e^{il\phi_p}$,
where $\alpha_l = (2^l l!)^{-1/2}u^l e^{-u^2/4}$.  Factoring the $\alpha_l$'s
out of the determinant leaves ${\bf S}_{\it R}$
in the form of an alternant, 
which may be readily evaluated, yielding 
\begin{eqnarray}
 \langle D, m_1,\dots,m_R \vert N,R,u;WN\rangle  =  \nonumber \\
C
\prod_{l=1}^R (m_l!)^{-1/2} 
\prod_{1\leq i<j \leq R} \sin \biggl(\frac{\pi}{R} (m_i-m_j)\biggr),
\end{eqnarray}
where $C$ is a constant common to each term in the expansion of a particular 
$\vert N,R,u;WN \rangle$.   Because of the factor 
$\prod_{i<j} \sin \biggl(\frac{\pi}{R} (m_i-m_j)\biggr)$, 
$\langle D, m_1,\dots,m_R \vert N,R,u;WN \rangle$ must vanish if 
$m_i \equiv m_j \bmod R$ for any pair $(m_i,m_j)$.  
This result, along with the observation that
there are exactly $R$ 
$m_i$'s, lead us to the following lemma: 
\begin{lemming}
{\rm 
For each integer $j$, $D \leq j \leq D+R-1$, each state 
$\vert D,m_1,\dots,m_R \rangle$ in the expansion of $\vert N,R,u;WN \rangle$ 
must include {\rm exactly} one single particle state $\psi_{\mu_j}$ 
(excluding states in the compact central disk) for which
$\mu_j \equiv j \bmod R$.  Specifically, we write $\mu_j -j = k_j R$, 
 where $k_j$ is a non-negative integer.
}
\end{lemming}

It follows that the total angular momentum of the states in the ring  
differs from the minimum possible value $\frac{1}{2}R(2D+R-1)$ by $kR$, 
where 
$k = \sum_{j=D}^{D+R-1} k_j$.
The significance of $k$ is discussed in the main body of the text. 
The lemma is very useful for computation, 
as it greatly reduces the number of terms in the expansion of a state
$\vert N,R,k;PN \rangle$, and allows 
us to project wavefunctions $\vert N,R,u;WN \rangle$ onto values of 
$L$ which would otherwise be inaccessible.


\begin{references}
\bibitem{previous} This has been confirmed by several previous studies, 
including Refs. 
\cite{lubin,chamon,macdonald,ferconi}.
\end{references}
\end{document}